# Hydrogen phosphate-mediated acellular biomineralisation within a dual crosslinked hyaluronic acid hydrogel

Ziyu Gao,[a,b] Layla Hassouneh,[a] Xuebin Yang,[a] Juan Pang,[c] Paul D. Thornton,*[,b] Giuseppe Tronci*[,a,d]

[a] Biomaterials and Tissue Engineering Research Group, School of Dentistry, St. James's University Hospital, University of Leeds, UK. E-mail: g.tronci@leeds.ac.uk.

[b] School of Chemistry, University of Leeds, Leeds, UK. E-mail: p.d.thornton@leeds.ac.uk.

[c] School of Material Engineering, Jinling Institute of Technology, Nanjing, China.

[d] Clothworkers' Centre for Textile Materials Innovation for Healthcare, School of Design, University of Leeds, UK.

**Graphical abstract**

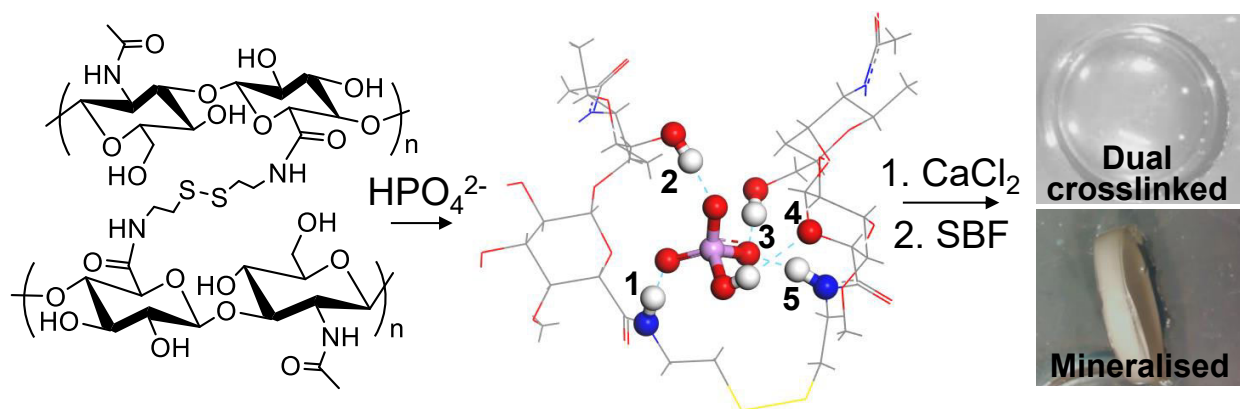

**Abstract**

The creation of hyaluronic acid (HA)-based materials as biomineralisation scaffolds for cost-effective hard tissue regenerative therapies remains a key biomedical challenge. A non-toxic and simple acellular method to generate specific hydrogen phosphate ($HPO_4^{2-}$) interactions within the polymer network of cystamine-crosslinked HA hydrogels is reported. Reinforced dual crosslinked hydrogel networks were accomplished after 4-week incubation in disodium phosphate-supplemented solutions that notably enabled the mineralisation of hydroxyapatite (HAp) crystals across the entire hydrogel structure. $HPO_4^{2-}$–cystamine-crosslinked HA hydrogen bond interactions were confirmed by attenuated total reflectance Fourier transform infrared spectroscopy (ATR-FTIR) and density functional theory (DFT) calculations. $HPO_4^{2-}$-mediated physical crosslinks proved to serve as a first nucleation step for acellular hydrogel mineralisation in simulated body fluid allowing HAp crystals to be detected by X-ray powder diffraction (2θ = 27°, 33° and 35°) and visualised with density gradient across the entire hydrogel network. On a cellular level, the presence of aggregated structures proved key to inducing ATDC 5 cell migration whilst no toxic response was observed after 3-week culture. This mild and facile ion-mediated stabilisation of HA-based hydrogels has significant potential for accelerated hard tissue repair *in vivo* and provides a new perspective in the design of dual crosslinked mechanically competent hydrogels.

**Keywords:** Hyaluronic acid; Hydrogel; Hydrogen phosphate interaction; Cystamine crosslinking; Biomineralisation.

**1. Introduction**

As one of the main components of extracellular matrix (ECM), hyaluronic acid (HA) has been applied widely in medicine, for example as a lubricant for osteoarthritis treatment [1] [2], wound dressing material to support healing [3] and as post-operation adhesive [4]. Recently, HA hydrogels have been developed as implants to support cell growth and aid regeneration of soft tissues including derm [3] [5], mucosa [4] [6] and tendon [7] [8], due to the biocompatibility, biodegradation profile and mechanical properties of HA. The advantageous features of HA in biology, as well as its chemical structure, which can be selectively targeted to fabricate mechanically competent bioinspired scaffolds, have also been leveraged to support the regeneration of bone. However, this has frequently required either severe or sophisticated synthetic approaches to address the mechanical and compositional requirements of bone. Although many methods have been investigated [9] [10], mild non-toxic routes enabling the fabrication of drug-free bone-like HA-based architectures have not yet been fully realized.

Ionic interactions, particularly salt effects, enable biomacromolecule crosslinking in a mild manner that avoids chemical synthesis and/or extensive energy radiation [11][12]. The Hofmeister effect details the extent that protein solubility is altered by the presence of different salts in an aqueous environment, and may be used to design protein-based hydrogels with enhanced compressive and tensile properties [13] [14]. The mechanism of salt effect on nonelectrolytes in aqueous solutions has been explored in (i) hydration theories, (ii) electrostatic theories, (iii) Van der Waals forces, and (iv) internal pressure concepts [15]. However, non-specific ion-mediated interactions may be applied universally in macromolecules [16][17][18]. Barrett hypothesized that a particular salt could act as either a stabilizer (i.e. kosmotrope) or a destabilizer (i.e. chaotrope) for a specific macromolecule [19]; for instance, alginate may be particularly well stabilized by calcium [20][21][22]. Leveraging aforementioned salt effects, we hypothesised that phosphate groups may enable the generation of additional physical crosslinks in a chemically crosslinked HA hydrogel network bearing amide net-points, on the one hand, and act as nucleation sites to accomplish hydrogel biomineralisation in near-physiologic conditions, on the other hand. Phosphate groups were selected as the most common component of buffer salts and since they are known to mediate protein denaturation [23], the stabilisation of HA-based electrospun fibres [24] and biomineralisation [25].

Despite the crucial role of HA in the ECM of biological tissues and the unique functions of phosphate groups in hard tissue repair, the interaction of phosphate ions with HA-based hydrogels has only partially been studied, suggesting limited control of molecular interactions and macroscopic effects [26]. Attempts to characterise the interaction between HA and the phosphate head group in phospholipid model membranes have been made through differential scanning calorimetry (DSC), fluorescence spectroscopy, small-angle X-ray scattering (SAXS), infrared spectroscopy (IR) and atomic force microscopy (AFM) [27]. However, the resulting phosphate ion-HA interaction was too insignificant to be observed by the above-mentioned methods. This underlines the experimental challenge in designing phosphate ion-mediated dual crosslinked HA-based hydrogel systems as a biomineralisation template for the direct build-up of bioinspired, mechanically competent HA matrices for hard tissue repair.

Other than phosphate-HA interactions, the integration of hybrid micromorphologies has attracted great interest in bone regeneration [28], and has been pursued in HA-based hydrogels aiming to realize bioinspired bone-like nanocomposites [10]. The *in situ* precipitation of calcium phosphate was reported on the surface of HA hydrogels, yielding a calcium phosphate nanocomposite on the outer layer of the hydrogel scaffold [9]. Ion diffusion methods have also been studied for mineralisation, including an electrophoresis approach [29] and a double-diffusion system [30]. However, only amorphous hydroxyapatite (HAp) was observed in the electrophoresis approach, whereas only calcium phosphate minerals were obtained via the sophisticated double-diffusion system. Consequently, accomplishing time-efficient and controllable formation of HAp crystals with native patterns and growing density is still a great challenge in the design of hierarchical 3-dimensional (3D) structures that mimic human bones [31]. Constructing a secondary crosslinked structure by including $HPO_4^{2-}$ in the hydrogel matrix may provide microchannels within the network that enable HAp formation, and consequently bone repair.

In this work, two HA-based hydrogels that contained either cystamine- or ethylenediamine-induced crosslinks were designed and assessed in phosphate-supplemented aqueous solutions and a range of salts that partially comprise the Hofmeister series, with the aim of developing a simple method to induce both dual crosslinking and HAp mineralisation across the hydrogel structure. We hypothesised that non-toxic phosphate-binding amide crosslinks could be introduced during the crosslinking reaction to control the swelling and mechanical properties of the HA-based hydrogels and lay down the foundation of a new bioinspired HA-based structure. The increased segment length of, and the presence of disulfide bridges in, cystamine-crosslinked (with respect to ethylenediamine-crosslinked) HA chains were hypothesised to minimise steric hindrance and enhance the yield of physical crosslinking and acellular biomineralisation during hydrogel incubation in phosphate-supplemented aqueous solutions. Incubation of the hydrogels in aqueous solutions supplemented with hydrogen phosphate ($HPO_4^{2-}$) generated hydrogen bonds acting as physical crosslinks, thereby yielding a very stable macrostructure with customisable mechanical properties. The mineralisation process of $HPO_4^{2-}$-conditioned HA hydrogels was monitored in conventional simulated body fluid (c-SBF), whereby unique hierarchical structure and gradients of HAp mineral were recorded across the entire hydrogel and confirmed by X-ray computed microtomography (μCT). The simplicity and mildness of this dual crosslinking and mineralisation approach enable method transferability to other biopolymers and offers great promise for the creation of drug-free bioinspired materials for cost-effective bone regenerative therapies.

## 2. Materials and methods

### 2.1. Materials

Hyaluronic acid sodium salt (molecular weight: 1,200 kDa, cosmetic grade) was purchased from Hollyberry Cosmetic, 4-(4,6-dimethoxy-1,3,5-triazin-2-yl)-4-methyl-morpholinium chloride (DMTMM) and 2-(*N*-morpholino) ethanesulfonic acid (MES) were purchased from Fluorochem. $(NH_4)_2SO_4$, $Na_2HPO_4 \cdot 7H_2O$, tris(hydroxymethyl)aminomethane (TRIS), cystamine dihydrochloride and ninhydrin reagent were purchased from Alfa Aesar. $Na_2SO_4$, $CH_3COONa$ (NaAc), $NaHCO_3$, $KCl$, $K_2HPO_4 \cdot 3H_2O$, $MgCl_2 \cdot 6H_2O$, $CaCl_2$, and ethylenediamine were ordered from VWR. Phosphate buffered saline (PBS) was purchased from Lonza. 2,4,6-trinitrobenzenesulfonic acid (TNBS), alamarBlue$^{TM}$ Cell Viability Reagent, CellTracker™ Green 5-chloromethylfluorescein diacetate (CMFDA) dye and the LIVE/DEAD$^{TM}$ cell stain kit were purchased from ThermoFisher Scientific. All other reagents were purchased from Sigma-Aldrich. Unless specified, all the general reagents were analytical grade.

### 2.2. Hydrogel preparation

HA hydrogels were fabricated according to our previous method [32]. HA powder was dissolved in MES buffer solution (0.1 M, pH 5.5) at room temperature in 2 wt.% concentration. DMTMM (2 equivalents per HA repeat unit) was then added at 37 °C to activate the carboxyl groups of HA. Following 1-hour activation at 37 °C, either cystamine or ethylenediamine was added with a molar ratio of 0.4 moles relative to the moles of each HA repeat unit. The stirring speed was increased to 1000 rpm for 5 minutes, and either 0.6 g or 0.8 g of the reacting solution was cast into 24-well plates. HA hydrogels were obtained after 2-hour incubation at 37 °C. Cystamine and ethylenediamine crosslinked HA hydrogels were named as C2-40 and E2-40, whereby C and E signify HA crosslinking with cystamine and ethylenediamine, respectively; 2 is the wt.% of HA in the hydrogel-forming solution, whilst 40 is the mol.% of each crosslinker added with respect to HA's carboxylic groups.

## 2.3. TNBS assay and determination of polymer crosslinking

Polymer crosslinking density was indirectly assessed via determining the concentration of unreacted amine groups presented by the crosslinkers (either cystamine or ethylenediamine) in each hydrogel using the 2,4,6-trinitrobenzene sulfonic acid (TNBS) assay [32]. 0.8 g of freshly synthesised hydrogel was freeze-dried without deionised water washing. Each dry network was immersed in 2 mL $NaHCO_3$ solution (4 wt.%) at 40 °C for 30 minutes to remove any unreacted cystamine or ethylenediamine. 1 mL of the supernatant was collected and incubated in dark (40 °C, 3 hours, 120 rpm) with 1 mL TNBS solution (0.5 wt.% in deionised water). 3 mL HCl (6 N) was added to the incubated solution, and the temperature raised to 60 °C for 1 hour to terminate the reaction. After cooling to room temperature, the sample solutions were diluted with 5 mL of deionised water. The unreacted TNBS was washed out by extraction with 20 mL diethyl ether (×3). 5 mL of the retrieved sample solution was incubated in hot water to evaporate any diethyl ether and diluted with 15 mL of deionised water. Finally, 2 mL of each solution was analysed by UV-Vis spectroscopy at 346 nm. Quantification of any cystamine or ethylenediamine residue was carried out by comparison with a cystamine or ethylenediamine calibration curve.

## 2.4. Hydrogel swelling tests

Various ion-hydrogel interactions were compared through changes in swelling ratio. Each replicate of prepared C2-40 and E2-40 hydrogels of known wet weight ($\omega_0$) was individually immersed in either $(NH_4)_2SO_4$, $Na_2SO_4$, $Na_2HPO_4·7H_2O$, $CH_3COONa$ (NaAc), NaCl or deionised water (50 mL solution). Swelling tests in PBS buffer solution (LONZA) and conventional simulated body fluid (c-SBF) were also carried out. The wet weight $(\omega_t)$ was recorded at different time points for up to 4 weeks. All the single-salt solutions used were prepared with 50 mM concentration and replaced by fresh solution every week with the same volume. The c-SBF solution was prepared as reported previously [33]. Briefly, all the salts were added to 960 mL deionised water in the following order: 8.036 g NaCl, 0.352 g $NaHCO_3$, 0.225 g KCl, 0.230 g $K_2HPO_4·3H_2O$, 0.311 g $MgCl_2·6H_2O$, 40 mL HCl (1.0 M), 0.293 g $CaCl_2$, 0.072 g $Na_2SO_4$, 6.063 g TRIS. The pH of the solution was buffered at pH 7.4 by adding HCl (1.0 M). The swelling ratio was calculated via eq. 1, as reported below:

$$Swelling\ ratio = \frac{\omega_t}{\omega_0} \times 100 \qquad (1)$$

**2.5. Hydrogel stability tests**

Hydrogels were incubated for 4 weeks in either the $Na_2HPO_4$-supplemented solution or deionised water. Following incubation, retrieved samples were washed by immersing in deionised water (×3) to remove any free salts and then freeze-dried. The relative mass of the hydrogel was calculated according to eq. 2 by measuring the dry weight of the freeze-dried freshly synthesized ($\omega_0$) and retrieved ($\omega_d$) samples, as reported below:

$$Relative\ mass = \frac{\omega_d}{\omega_0} \times 100 \qquad (2)$$

**2.6. Hydrogel compression tests**

Hydrogel compression properties were measured using a Bose ELF 3200 apparatus with a 0.02 mm/s compressive rate. All replicates were cut into 3 mm diameter cylinders. Compression stress and strain of either initial or salt-treated C2-40 and E2-40 hydrogels were evaluated and compared.

**2.7. Morphology study of the hydrogel network following salt treatment**

Hydrogel morphology was observed using a HITACHI 3400 scanning electron microscope (SEM) under 20 kV voltage with gold coating. All hydrogels were treated with different salts for 4 weeks and flushed with deionised water. SEM analysis was carried out on freeze-dried hydrogel networks. Samples were carefully transferred into 6-well cell culture plates and frozen at -20 °C prior to lyophilisation, to minimise lyophilisation-induced sample shrinking. During the course of incubation, the hydrogel structures were also observed by optical microscopy (Zeiss) at different time points after various treatments.

**2.8. Mechanistic study**

$HPO_4^{2-}$ interaction with the HA hydrogels was investigated via attenuated total reflectance Fourier transform infrared spectroscopy (ATR-FTIR, Bruker spectrometer) at room temperature and density functional theory (DFT) calculations. The optimised structures were obtained by DFT calculations at b3lyp/6-31G(d) level carried out using Gaussian 16 program [34]. The binding energy between the HA repeat unit and $HPO_4^{2-}$ was simply calculated as [$\Delta E = E_{total} - (E_{HA} + E_{HPO4})$], in which the single point energy was calculated at b3lyp/6-311+G(d,p) level. For display, blue dashed lines indicated the hydrogen bonds, oxygen (O) atoms were depicted in red, nitrogen (N) in blue, sulfur (S) in yellow, carbon (C) in grey, hydrogen (H) in white and phosphorus (P) in pink. All the atoms which were involved in hydrogen bond formation are depicted as spheres.

**2.9. Cell adhesion study**

ATDC 5 chondrocytes (chondrogenic cell line) were used as non-mineralising joint resident cells of the bone-cartilage interface. The initial C2-40 network (which was proven to mediate secondary interactions with phosphate ions) was washed by sterile deionised water (×3) and basal cell culture medium (BM) (×3). BM was composed of 50 vol.% Dulbecco's modified eagle's medium (DMEM, D6546) and 50 vol.% Ham's nutrient mixture F12 (12-615), and supplemented by 5 % fetal calf serum (FCS) and 1% penicillin and streptomycin (PS). The final concentration of phosphorus in BM was 0.884 mM. Cells were labelled by CellTracker™ Green (CMFDA) and re-suspended in medium with a cell density of $2×10^5$ cells/mL. 100 μL cell suspension ($2×10^4$ cells) was injected on the surface of each hydrogel (n=3). 2 mL of BM was added into each well after 3 hours seeding. Cell attachment and growth was observed and recorded after 48 hours by fluorescence/optical microscopy (Zeiss).

To study the influence of $Na_2HPO_4$ on cell migration, BM was replaced by $Na_2HPO_4$ treated medium (TM) after 1-week of culture. The cell culture in TM was named as "conditional cell culture" and this culture time started when the medium was replaced. TM was prepared from the basal medium via supplementation of sterile $Na_2HPO_4·7H_2O$ powder to achieve a final concentration of 1.884 mM (1 mM increase in phosphate compared with BM). In the control group, the medium was replaced by fresh BM. ATDC 5 cell attachment and growth were investigated by fluorescence/optical microscopy (Zeiss), cell migration was studied via Laser scanning confocal microscopy (LEICA TCS SP8, excitation wavelength 488 nm). All the samples were washed by sterile PBS (×3) to remove any dead cells and impurities before calcein-AM staining.

## 2.10. Acellular mineralisation

Hydrogel C2-40 was selected for the biomineralisation study given its capability to mediate secondary interactions with phosphate ions (confirmed by swelling and compression measurements). After 4-week immersion in $Na_2HPO_4$ solution (50 mM, 1.0 L, 37 °C), C2-40 hydrogels (n=3) were transferred into an excess of deionised water for 24 hours to remove any free phosphate salt, whereby the deionised water was replaced for three times during this time period. Washed hydrogels were then soaked in 200 mL calcium chloride (10 mM) for another 24 hours [35]. Calcium-treated C2-40 samples were flushed by deionised water to remove any surface salt and subsequently soaked in 1.0 L c-SBF for mineralisation at 37 °C for 2 weeks. Non-$Na_2HPO_4$-treated C2-40 samples were immersed in $CaCl_2$ (10 mM, 200 mL) for 24 hours and underwent the same mineralisation procedure as a control group. The mineral structure was confirmed by X-Ray powder diffraction (XRD) at room temperature in the range of 2q of 20°-60°. Freeze-dried initial and mineralised C2-40 networks, as well as mineralised C2-40 networks after being burnt at 1000 °C for 30 minutes, were measured. The 3D structure of mineralisation was investigated by X-ray computed microtomography (µCT) (Skyscan 1072, Bruker, Kontich, Belgium). Samples were scanned at 100 kVp, 100 mA, and 11.19 µm pixels, with a 1-mm aluminium plus copper filter and a scanning time of around 60 minutes. A reconstruction software program (NRecon; SkyScan) was used to convert the raw data into bitmap (bmp) files. 3D alignment and registration of samples were done using Data Viewer software (v1.4.3; Bruker microCT). Both CTan and CTvol (v1.10.11.0; Bruker microCT) software were used for the 3D structural analysis.

## 2.11. Statistical analysis

All the results were analysed with at least three replicates (n≥3). The results are presented as mean±SD. The significant difference was calculated through One-way ANOVA analysis with a p-value at 0.05, which was considered as significant. These were labelled as *p < 0.05, **p < 0.01, ***p < 0.001, ****p < 0.0001.

## 3. Results and discussion

### 3.1. Hydrogel crosslinker density

The crosslinker density was evaluated by the TNBS assay (Figure S1, Supp. Inf.) to determine the quantity of ethylenediamine or cystamine included within the covalent network formed [32]. When adding 40 mol.% of either cystamine or ethylenediamine, approximately 25 mol.% of crosslinker reacted with HA during gel formation (**Table 1**), ensuring that a comparable crosslink density was accomplished in both hydrogel networks regardless of the crosslinker used.

**Table 1.** Composition of HA hydrogels crosslinked with either cystamine (C2-40) or ethylenediamine (E2-40). TNBS assay was employed to quantify the crosslinker quantity in the HA network. Results are presented as Mean±SD.

| Sample ID | HA concentration (wt.%) | Crosslinker quantity (mol.% of -COOH) | |
|---|---|---|---|
| | | Added in | Reacted |
| C2-40 | 2.0 | 40.0 | 25.30±0.85 |
| E2-40 | 2.0 | 40.0 | 25.27±0.01 |

### 3.2. Swelling behaviour of HA hydrogels

Hydrogel swelling equilibrium was reached after 1 day for both C2-40 and E2-40 hydrogels following incubation in single salt-supplemented solutions (**Figure 1, a&b**). The swelling ratio (SR) of C2-40 samples was found to be in the region of 150 wt.% in all salt solutions, whilst a swelling ratio of 425 wt.% was measured in deionised water. The interaction of selected salts with C2-40 resulted in decreased hydrogel swelling, following the order $Na_2HPO_4$ > $(NH_4)_2SO_4$ = $Na_2SO_4$ > NaCl > NaAc. For sample E2-40, lower swelling ratio values (~325 wt.%) were observed in deionised water, compared with hydrogel C2-40. However, anion-HA hydrogel interactions were obvious in the 4-week swelling study and followed the order $Na_2SO_4$ ≥ $(NH_4)_2SO_4$ > NaCl > NaAc > $Na_2HPO_4$. When comparing the two hydrogels, the most striking difference is observed in the swelling behaviour in $Na_2HPO_4$-supplemented solutions. Following incubation of sample

C2-40 in $Na_2HPO_4$, a significant decrease of SR was observed from 123±2 wt.% (1 day) to 62±1 wt.% (28 days). In contrast, the SR of E2-40 in $Na_2HPO_4$ increased from 140±5 wt.% (1 day) to 161±4 wt.% (28 days). This observation was hypothesised to reflect the specific hydrogen bond interaction between $HPO_4^{2-}$ and the cystamine-crosslinked HA chains.

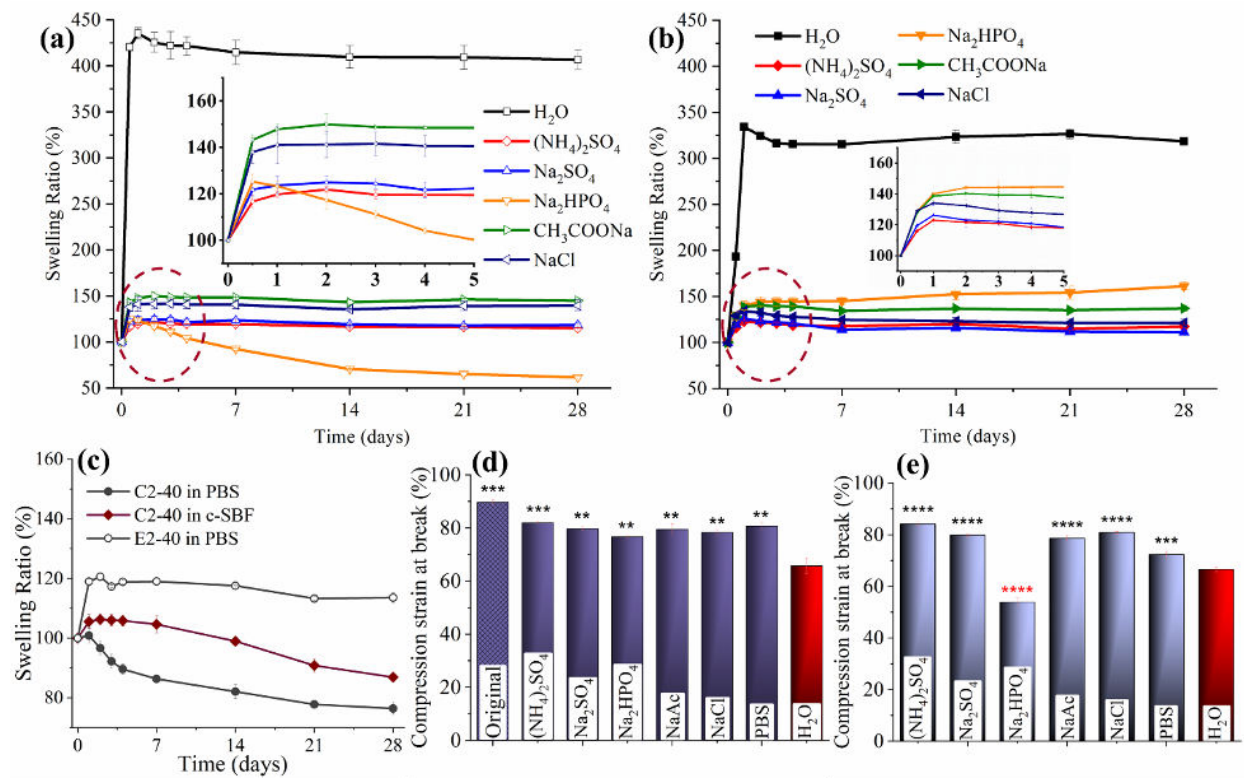

**Figure 1.** Effect of salt-supplemented solution on hydrogel swelling and compressive properties. (a-b): Swelling ratio of C2-40 (a) and E2-40 (b) hydrogels in $(NH_4)_2SO_4$, $Na_2SO_4$, $Na_2HPO_4$, NaAc, NaCl and deionised water ($H_2O$). Insert graphs: Swelling Ratio (%, Y axis) profile over time (days, X axis). (c): Swelling ratio of C2-40 and E2-40 hydrogels in PBS and c-SBF. (d-e): Compression strain at break measured with hydrogel C2-40 following synthesis ('Original', surface flushed by deionised water before testing) and either 1-day (d) or 4-week (e) incubation in single salt-supplemented solutions. Statistical analysis is presented with respect to the $H_2O$ group and labelled as **$p < 0.01$, ***$p < 0.001$, ****$p < 0.0001$. Data are presented as Mean±SD.

As observed in the $Na_2HPO_4$-supplemented solution, the swelling ratio of samples C2-40 in PBS solution (**Figure 1c**) presented a similar decreasing trend over time (SR: 101±3 wt.% (1 day) → 76±2 wt.% (28 days)), supporting the hypothesis that phosphate ions lead to a reduction in hydrogel swelling. However, the swelling ratio of ethylenediamine crosslinked hydrogel (E2-40) was stable (~120 %) for the first 7 days before marginally decreasing over the next 21 days (SR: 119±1 wt.% (7 days) → 114±2 wt.% (28 days)) (**Figure 1c**). Based on the significant decrease in SR measured in hydrogel C2-40 following incubation in both $Na_2HPO_4$ and PBS solution, the swelling ratio was also recorded in c-SBF to further elucidate any $HPO_4^{2-}$-mediated interaction with cystamine-crosslinked HA. As expected, a similar but slower decrease in hydrogel swelling was recorded in c-SBF over time, which is likely due to the different phosphate concentrations across the selected solutions (**Table 2**).

**Table 2.** Swelling ratio of C2-40 hydrogels following 4-week incubation in phosphate-supplemented solutions. The results are presented as Mean±SD.

| Solution Name | Phosphate concentration (mM)* | Swelling ratio (%) |
| --- | --- | --- |
| $Na_2HPO_4$ | 50 | 62±1 |
| PBS | 6.658 | 76±2 |
| c-SBF | 1.001 | 87±1 |

*Concentration of hydrogen phosphate and dihydrogen phosphate.*

The significant difference in swelling ratio of hydrogel C2-40 was therefore attributed to the interactions between $HPO_4^{2-}$ ions and cystamine-crosslinked hyaluronic acid, offering a new dimension for adjusting the swelling of the hydrogel by altering the chemical composition of the crosslinker.

Other than the swelling behaviour, the stability of C2-40 hydrogel was determined by quantifying its relative mass following 4-week incubation in the $Na_2HPO_4$-supplemented aqueous solution (Figure S2, Supp. Inf.). Although a decrease in mass was observed in $Na_2HPO_4$-treated C2-40 networks with respect to deionised water-treated controls, a relative mass of 67.7±1.8 wt.% was measured compared to 71.7±0.4 wt.% for the hydrogel controls, verifying good material stability and limited $Na_2HPO_4$ impact.

### 3.3. Compressive properties of salt-treated hydrogels

Both C2-40 and E2-40 hydrogels were reinforced by ions to some degree, whereby stiffer networks and varying values of compression strain (**Figure 1, d&e**) and stress at break (Table S1) were measured, which further proved the effect of salts on hydrogel mechanical properties. Among the salt-treated samples, the most interesting phenomenon was observed in C2-40 hydrogels incubated in the $Na_2HPO_4$ environment, whereby the lowest value of compression strain at break (77±0.3 %) was recorded after 1 day before decreasing to 54±1.5 % after 4-week treatment. All the other groups formed a relatively stable network (**Figure 1, d&e**). This observation further supported the development of selective, strong $HPO_4^{2-}$-mediated physical crosslinks in the C2-40 hydrogel following salt treatment, so that the mechanical behaviour of the resulting dual crosslinked hydrogel network could be adjusted from elastic to stiff. This variation in mechanical behaviour was also supported by the trends of compression stress at break measured in $Na_2HPO_4$-treated and water-incubated groups after 1-day and 4-week treatment (Figure S3&S4). The additional interactions between $HPO_4^{2-}$ groups and the cystamine-crosslinked HA network were therefore investigated as a means to induce acellular biomineralisation of cystamine-crosslinked HA hydrogel.

### 3.4. Morphology of salt-treated hydrogels

To study whether hydrogel surface morphology is affected by the salt treatment, freeze-dried hydrogels were inspected by SEM after 4-week incubation in $Na_2HPO_4$-supplemented solution. Crystal-like salts were not observed by SEM in both samples C2-40 (**Figure 2**) and E2-40 (Figure S5), suggesting that salts diffused into the hydrogel and attached to the network, in agreement with the salt-enhanced compression properties and decreased swelling ratio (**Figure 1**).

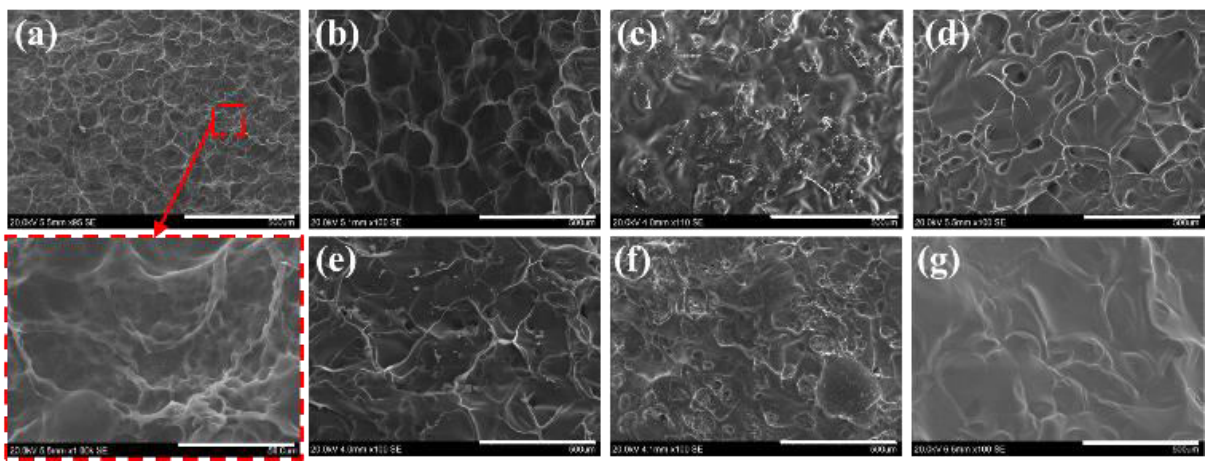

**Figure 2.** SEM images of freeze-dried C2-40 networks following 4-week incubation in aqueous solutions. (a): $(NH_4)_2SO_4$ (including zoomed-in image below); (b): $Na_2SO_4$; (c): $Na_2HPO_4$; (d): NaAc; (e): NaCl; (f): PBS; (g): $H_2O$. Scale bar of (a-g): 500 μm. Scale bar of zoomed-in image of (a): 50 μm.

All retrieved samples exhibited comparable porous-like surfaces, indicating minimal impact of the incubation process with either salt-supplemented incubating media (Figure 2 a-f) or salt-free deionised water (Figure 2 g). This, together with the non-detection of crystal-like salts on the hydrogel surface, suggests that any interaction of salt species, i.e. phosphate groups, with HA's covalent network occurred at the molecular rather than microscopic scale.

To further elucidate the extent of the above-mentioned ion interactions, hydrogels were incubated for three weeks in the presence of $Na_2HPO_4$ (50 mM), PBS solution and c-SBF. Aggregation of the hydrogel surface was observed in retrieved samples (**Figure 3**) after 3-week treatment in either $Na_2HPO_4$ (**Figure 3a**) or c-SBF (**Figure 3c**), whilst no visible effect was seen in hydrogels incubated for three weeks in PBS. On the other hand, when the incubation time in PBS solution was extended from 3 weeks to 3 months, aggregated structures with regular gaps were clearly visible in C2-40 hydrogels (**Figure 3 d&e**), as highlighted by the red arrows.

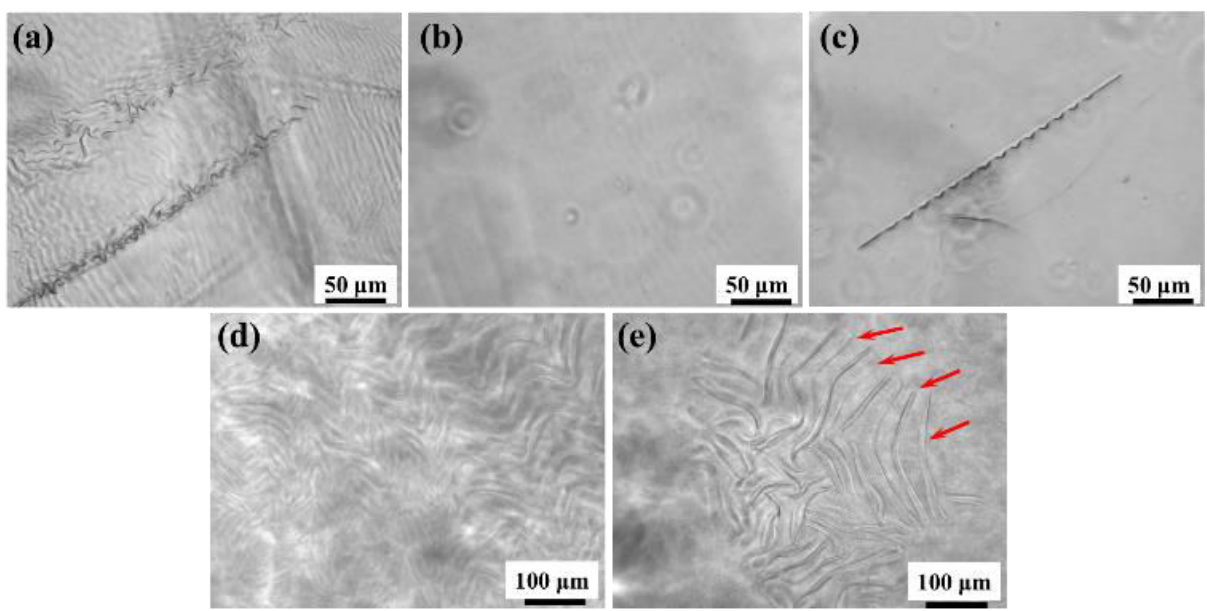

**Figure 3.** Optical images of hydrogel C2-40 following either 3-week incubation in $Na_2HPO_4$ (a), PBS (b), and c-SBF (c), or 3-month incubation in PBS buffer (d and e). Deep aggregation of the network was indicated with red arrows.

Given the absence of crystal-like aggregates via previous SEM analysis, aforementioned microscale effects are likely attributed to the development of strong interactions between hydrogel C2-40 and $HPO_4^{2-}$ ions, whereby the decreased yield of aggregation in PBS with respect to $Na_2HPO_4$ is attributed to the slow formation of hydrogen bonds and the decreased concentration of phosphate ions (**Table 2**) in the former compared to the latter medium. This aggregation mechanism provided the opportunity to create reinforced dual crosslinked hydrogel networks in near-physiological conditions (as indicated by previous compression tests in **Figure 1 d&e**) and an easy and stable method to build up $HPO_4^{2-}$ nucleation sites in the hydrogel for subsequent acellular biomineralisation.

### 3.5. Mechanistic study of $HPO_4^{2-}$-induced physical crosslinking

The development of physical crosslinks between $HPO_4^{2-}$ and cystamine-crosslinked HA was further supported by density functional theory (DFT) calculations. Three models of cystamine-crosslinked HA (C-HA) were optimised according to their energy minimum configuration. As presented in **Figure 4a**, the most stable structure was achieved in model C-HA3 ($\Delta E_{C-HA3}$= -170.751 kcal/mol), whilst increased total interaction energies were measured with the other two models ($\Delta E_{C-HA1}$= -162.075 kcal/mol, $\Delta E_{C-HA2}$= -169.501 kcal/mol). In the most stable model C-HA3, three atoms of oxygen (O) in the $HPO_4^{2-}$ species engages in hydrogen bonds with the NH (1, 5) and OH (2, 3) groups of crosslinked HA, whilst the OH group in $HPO_4^{2-}$ forms hydrogen bonds with the O atom of HA (4).

To investigate the influence of both the disulfide bridge and the number of carbon atoms in the crosslinking chain, the same binding sites as in C-HA were calculated in HA structure models crosslinked with either 1,6-hexanediamine (6 carbon atoms), butane-1,4-diamine (4 carbon atoms) or ethylenediamine (2 carbon atoms), and abbreviated as H-HA, B-HA, E-HA, respectively. As presented in **Figure 4b**, the strongest interaction in the H-HA structure was obtained in model H-HA3 with a $\Delta E_{H-HA3}$= -162.149 kcal/mol (Figure S6 and Table S2), which was 8.602 kcal/mol lower than the one recorded in model C-HA3 ($\Delta E_{C-HA3}$= -170.751 kcal/mol). Although no direct binding contribution of the S-S bridge was observed, the optimised structure and the reduced binding energy proved an indirect effect. In B-HA models, a $\Delta E_{B-HA3}$ of -167.491 kcal/mol was calculated in the most stable configuration, hinting at a lower interaction compared to the model of 1,6-hexanediamine-crosslinked HA. Since butane-1,4-diamine is two carbon atoms shorter than 1,6-hexanediamine, the lower interaction measured in model B-HA3 with respect to H-HA3 suggests that the crosslinker length affects the development of $HPO_4^{2-}$- mediated physical crosslinks in the HA crosslinked chain. This observation is supported by the energy calculations in model E-HA, describing HA chains crosslinked with ethylenediamine as the shortest crosslinker of the three. Only one stabilised structure was obtained in this work, with a final $\Delta E_{E-HA}$ of -155.330 kcal/mol. Nevertheless, the lack of stable configurations of E-HA is against the development of $HPO_4^{2-}$-mediated hydrogen bonds in ethylenediamine-crosslinked HA, thereby supporting the role of the crosslinker length in the development of phosphate ion-HA secondary structures.

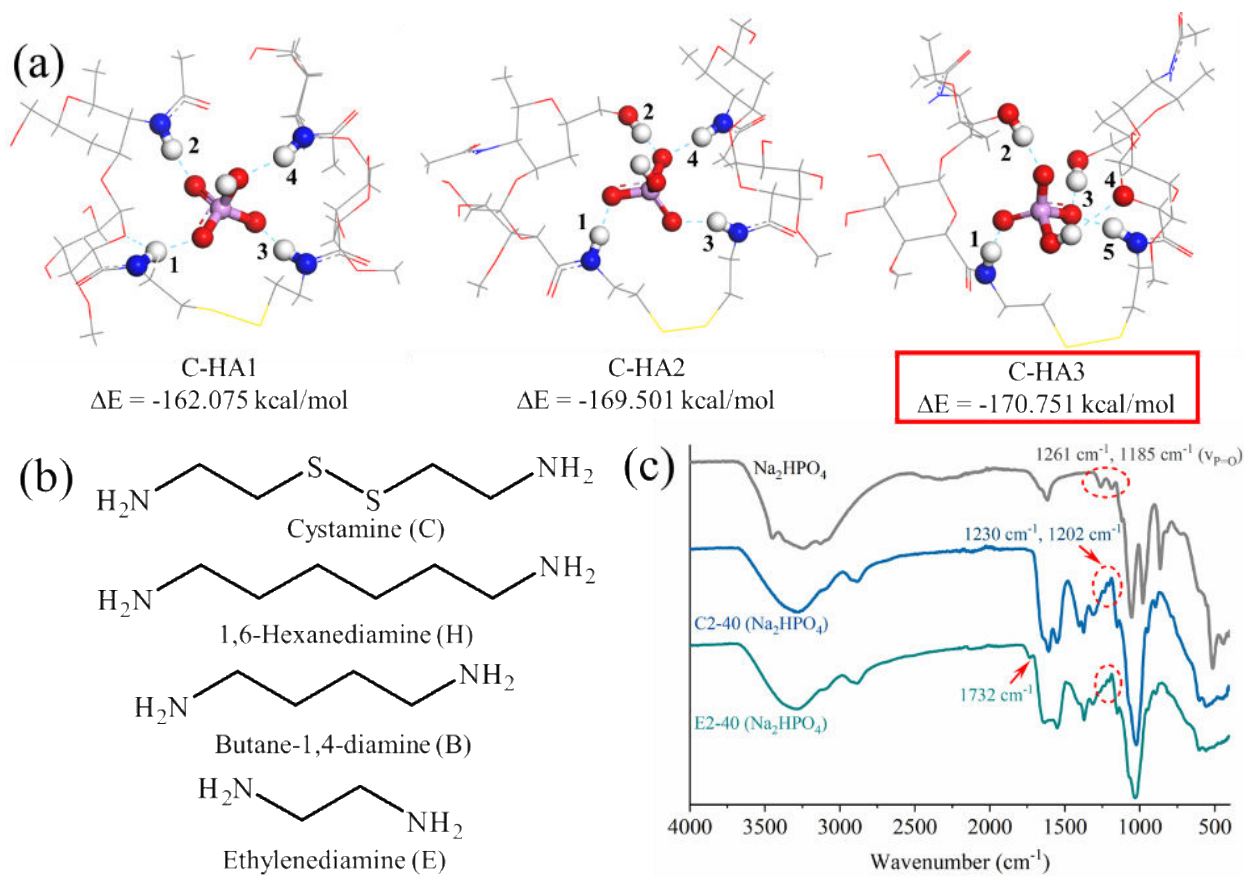

**Figure 4.** (a) DFT calculations of the hydrogen bond interaction between $HPO_4^{2-}$ ions and cystamine-crosslinked hyaluronic acid (C-HA1, C-HA2 and C-HA3). In all models, oxygen (O) atoms were presented in red, nitrogen (N) in blue, sulfur (S) in yellow, carbon (C) in grey, hydrogen (H) in white and phosphorus (P) in pink. (b) Molecular structure of computed crosslinkers. (c) IR spectrum of $Na_2HPO_4$ (top), $Na_2HPO_4$-treated C2-40 network (middle) and E2-40 network (bottom) following 4-week treatment.

Experimentally, a band corresponding to a P=O vibration was observed in the IR spectrum of the $Na_2HPO_4$-treated networks. New peaks at 1230 cm$^{-1}$ and 1202 cm$^{-1}$ were displayed by both C2-40 and E2-40 samples, which reflect the 1261 cm$^{-1}$ and 1185 cm$^{-1}$ peaks of $Na_2HPO_4$ (**Figure**

**4c**). C2-40 and E2-40 hydrogels were washed with deionised water for 24 hours to remove any free $Na_2HPO_4$ residue and freeze-dried prior to IR measurement. The existence of a shifted peak related to the P=O vibration provided strong evidence for hydrogen bond formation between P=O and cystamine-crosslinked HA units. The most interesting phenomenon was the almost disappearance of the original 1700 cm$^{-1}$ peak in the IR spectrum of the $Na_2HPO_4$-treated sample C2-40 (Figure S7), which is attributed to the amide linkage of HA (position 5, **Figure 4a**) [32] and which is still clearly visible in the IR spectrum of sample E2-40 following the same salt treatment. The hydrogen bond between the $HPO_4^{2-}$ ion and the nitrogen atom (N) of the amide bond (position 5, **Figure 4a**) may shift this peak to 1640 cm$^{-1}$. This result strongly supports the mechanism of multiple hydrogen bonds formed between $HPO_4^{2-}$ ions and the cystamine-crosslinked HA chains.

As the most stable interaction was obtained when the phosphate-amide site binding occurred, in agreement with Barrett's work on hyaluronic acid solutions [19], we propose that minimising steric hindrance by adjusting the length of the crosslinker is critical to providing proper access to $HPO_4^{2-}$ ions and enabling coordination and physical crosslinking with amide bonds. Furthermore, the introduction of disulfide bridges in the HA network provided HA-crosslinked chains with increased flexibility and increased opportunities for developing secondary interactions with phosphate groups [36]. This potential intermolecular interaction may induce the rearrangement of the disulfide bonds and hydrophilic-hydrophobic sites so that detectable effects can be observed at the macroscale and influence the material properties as shown in our results.

### 3.6. Cell adhesion study during $HPO_4^{2-}$ treatment

Following the results obtained in acellular conditions, an *in vitro* study was carried out with ATDC 5 chondrocytes. Chondrocytes were selected as non-mineralizing joint-resident cells, aiming to investigate both the material-induced cell response and any cell culture-induced effect on the material morphology. After 2-day cell culture in basal medium, some aggregated HA network was already observed on the surface of freshly synthesised hydrogel C2-40 (Figure S8), in line with the presence of phosphate groups in the cell culture medium (0.844 mM in BM). The aggregation kinetics were accelerated with respect to previously discussed acellular conditions, an observation which can be explained by considering the multiple ingredients in cell culture medium and cell metabolism. At the cellular level, the fluorescently labelled live cells aligning along the aggregated structure are visible (Figure S8), whereby the weak fluorescence is likely due to the quenching of the cell-labelling dye following cell growth.

After 1-week cell attachment and migration in basal medium, the conditional cell culture was carried out by replacing the medium with either fresh TM (1.884 mM phosphate) as testing group or BM (0.884 mM phosphate) as the control group. After 3 days of conditional cell culture and consequent calcein-AM staining, few fluorescent cells were observed via 3D confocal microscopy in either the BM or the TM group (**Figure 5**; higher resolution images are available in Figure S12 and Figure S13, Supporting Information).

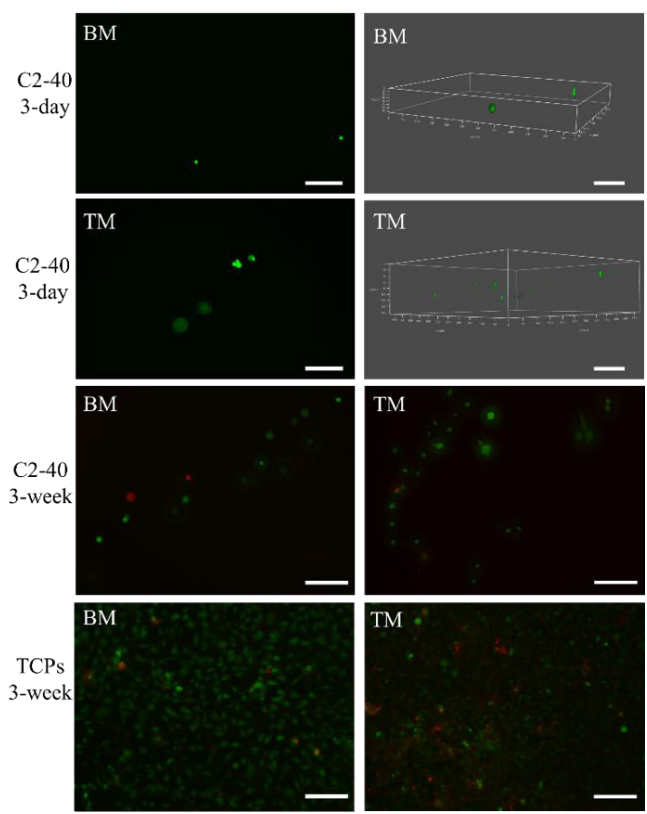

**Figure 5.** Conditional culture of ATDC 5 cells. Cells adhesion study on the surface of C2-40 hydrogels in either basal medium (BM) or $Na_2HPO_4$ treated medium (TM) after 3 days (first and second row). Cells after 3-week conditional culture on C2-40 hydrogel surface (third row) and TCPs (bottom row) in either BM or TM group. Live labelling was presented in green and dead labelling was indicated in red. Scale bar: 100 μm.

To confirm this, C2-40 hydrogels without cells were set as a blank control, whereby only one fluorescent dot with a maximum length of 10 μm was observed in the confocal image (Figure S9). This observation is unlikely to be related to living cells and is mostly attributed to impurity or fluorescence from $HPO_4^{2-}$ aggregation, as the cells observed in the hydrogels were approximately 30 μm in length and 10 μm in width (**Figure 5, first and second row**; higher resolution images are available in Figure S12, Supporting Information). When ATDC 5 cells were independently seeded on the surface of each initial C2-40 hydrogel in both BM and TM, most of the cells were found to adhere to the tissue culture plates (TCPs) rather than attach to the hydrogel networks. This observation suggests that a tighter network may help to minimise cell attachment and reduce the rate of degradation [37].

Live&dead staining results proved that $Na_2HPO_4$ treatment was non-toxic during a 3-week conditional cell culture period after comparing with BM groups, regardless of the hydrogel or TCP surface (**Figure 5, third and bottom row**; higher resolution images are available in Figure S13, Supporting Information). These results demonstrate that the $HPO_4^{2-}$-induced hydrogel aggregation process provides a regular channel for cell attachment and growth on the HA hydrogel surface (**Figure 5, TM group**), unlike the freshly synthesised C2-40 hydrogel. The acellular fabrication of previously described salt-mediated microstructures (**Figure 3**) can be proposed as the first stage of the hydrogel biomineralisation process. Here, the absence of cells is key to minimizing the risk of cell aggregation on the hydrogel surface, which could otherwise induce steric effects and delay HAp crystallisation.

### 3.7. Characterisation of HAp growth within the HA-based hydrogel

Both $Na_2HPO_4$ and non-$Na_2HPO_4$-treated C2-40 replicates were transferred into deionised water for 24 hours to remove any free salt, and then further treated with 200 mL $CaCl_2$ (10 mM) for 24 hours. The calcium concentration was chosen from a study on milk as one of the main sources for calcium supplementation [35]. All the samples were flushed with deionised water before the mineralization process, which was subsequently carried out in 1.0 L c-SBF at 37 °C. Remarkably, a homogeneous HAp phase was formed in the HA network C2-40 (**Figure 6a, left**), with full mineralisation observed across the whole hydrogel structure (Figure S10). No visible mineral was observed in the control group obtained without $HPO_4^{2-}$ treatment, ensuring that the hydrogel surface remained transparent (**Figure 6a, right**). To further characterise the mineralised structure of $Na_2HPO_4$-conditioned hydrogel, XRD diffraction was carried out (**Figure 6b**). A clear stacking structure corresponding to HAp was observed after burning the sample at 1000 °C for 30 minutes.

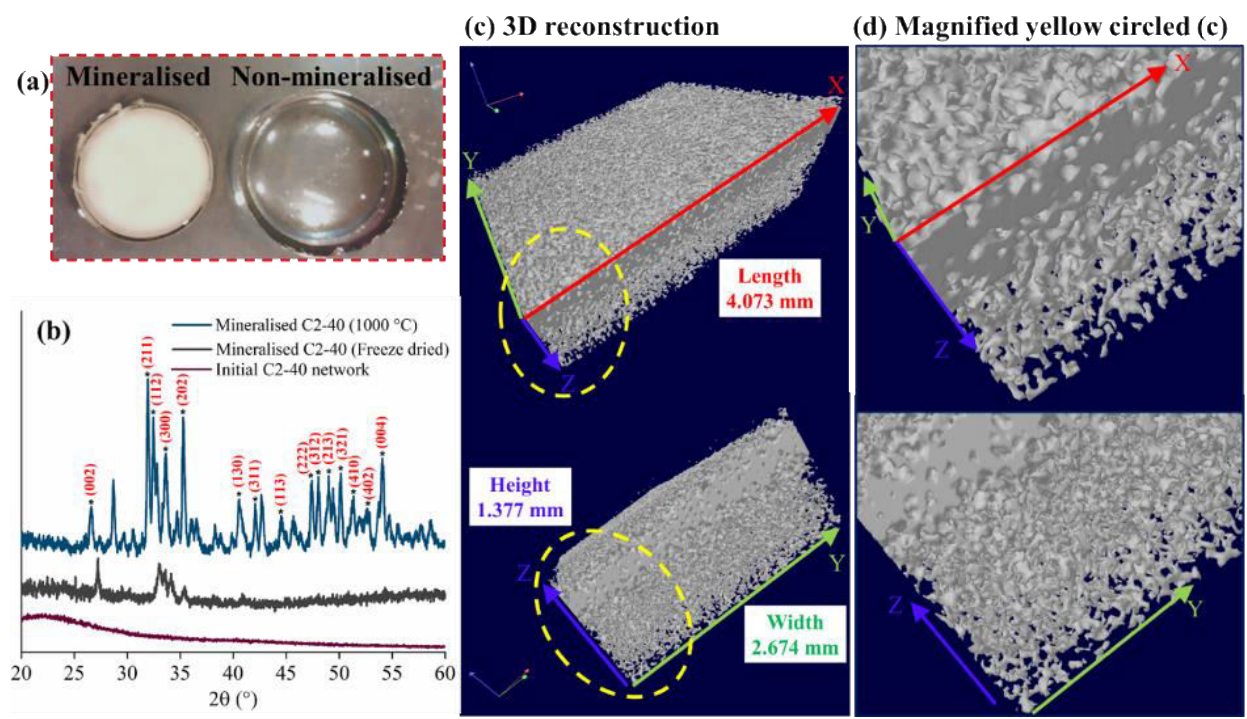

**Figure 6.** (a) Optical graphs captured at the end of the c-SBF incubation with both the $HPO_4^{2-}$-treated hydrogel C2-40 (left) and the corresponding $HPO_4^{2-}$-free hydrogel control C2-40 (right). (b) XRD spectra of the mineralised C2-40 after burning at 1000 °C (top, blue), the freeze-dried mineralized network (middle, black) and the initial freeze-dried C2-40 network (bottom, purple). (c-d) 3D reconstruction of the mineralised hydrogel C2-40: length-height direction (top) and height-width direction (bottom).

Some diffraction was recorded for the freeze-dried network at 2θ = 27°, 33° and 35°, again corresponding to the HAp phase [38], whilst no peak was observed in the initial C2-40 network. In addition to XRD spectra and digital macrographs, μCT was carried out as a non-damaging technique to visualise the 3D macrostructure of the mineralised C2-40 composite obtained following 2-week incubation in c-SBF (Figure S11). The cross-sectional image clearly reveals the decreasing HAp density from the top to the bottom side of the sample, in agreement with the results obtained from the μCT 3D reconstructed models (**Figure 6c&d**), and in contrast to the ion distribution surrounding the gel surface or limited formation of minerals [9][29]. This result demonstrates the high potential of the HAp-mineralized C2-40 hydrogel as a scaffold for hard tissue repair, particularly as gradient hydrogels for tissue regeneration [39].

## 4. Conclusions

The effect of the inclusion of a range of salts within cystamine and ethylenediamine-crosslinked HA-based hydrogels was investigated to prepare dual crosslinked bioinspired bone-like nanocomposites. Specific and strong hydrogen bond interactions acting as physical crosslinks were first discovered between cystamine-crosslinked HA chains and $HPO_4^{2-}$ groups, as indicated by the decreased swelling and decreased compression at the break, IR spectroscopy and DFT calculations. The introduction of phosphorus nuclei was key to enabling this interaction, which was successfully leveraged to accomplish HAp growth across the entire hydrogel structure. Gradient HAp structures were obtained and visualised by µCT 3D reconstruction after hydrogel incubation in 1.0 L c-SBF for 2 weeks. A novel method to generate dual crosslinked and mineralised structures is reported that is potentially significant for hard tissue repair.

**Acknowledgements**

The authors gratefully acknowledge Dr. Dong Xia for the XRD analysis and Dr. Sarah Myers for technical assistance.

**Declaration of competing interests**

The Authors declare no competing financial interests.

**Funding**

This research did not receive any specific grant from funding agencies in the public, commercial, or not-for-profit sectors.

**Supporting information**

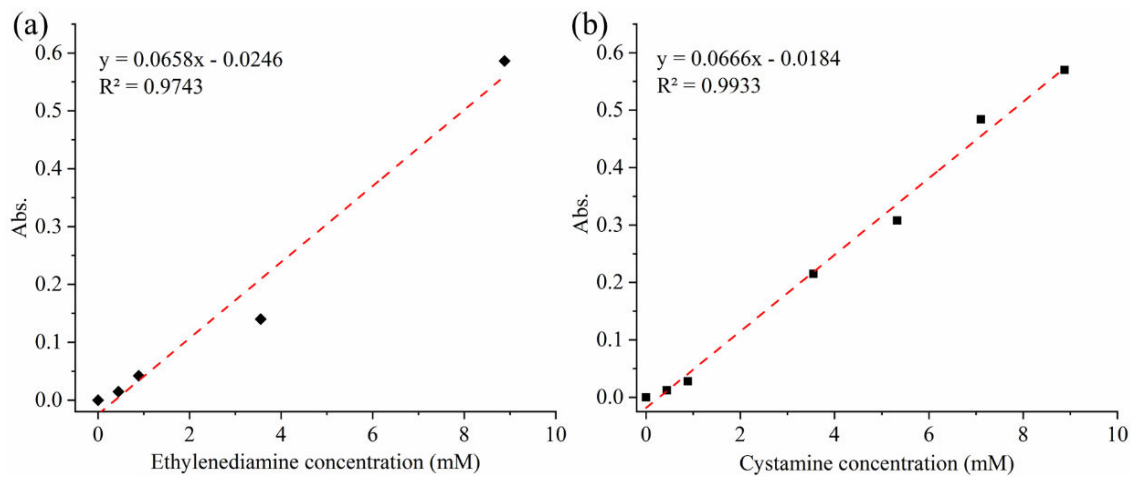

**Figure S1.** TNBS calibration curve of ethylenediamine (a) and cystamine (b)

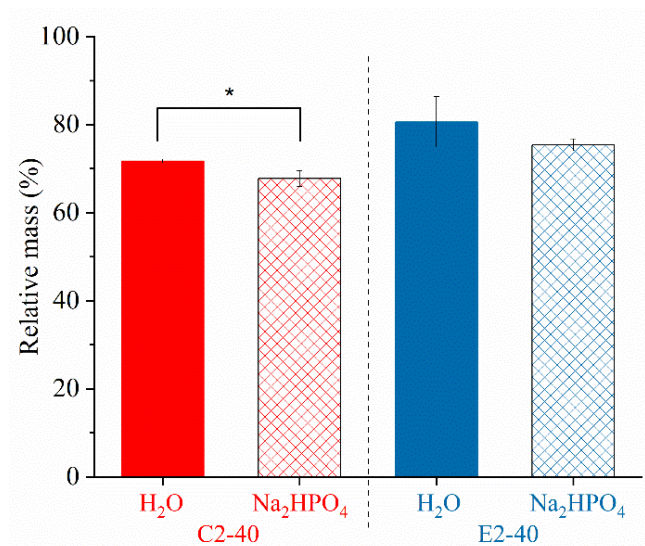

**Figure S2.** Relative mass of hydrogels (n=4) C2-40 (red) and E240 (blue) after 4-week immersion in either the Na2HPO4-supplemented solution (50 mM) or deionised water.

**Table S1.** Compression stress and strain values at break of C2-40 and E2-40 hydrogels after synthesis ('Original') and different salt treatments.

| Treating conditions | | C2-40 hydrogels | | E2-40 hydrogels | |
|---|---|---|---|---|---|
| | | Stress at break (kPa) | Strain at break (%) | Stress at break (kPa) | Strain at break (%) |
| 1 day | Original | 425±20 (***) | 90±1.0 (***) | 211±18 (****) | 77±0.8 (****) |
| | $(NH_4)_2SO_4$ | 227±48 (*) | 82±0.5 (***) | 146±15 (***) | 77±0.9 (****) |
| | $Na_2SO_4$ | 199±74 (-) | 80±0.8 (**) | 132±29 (**) | 78±0.4 (****) |
| | $Na_2HPO_4$ | 167±18 (*) | **77±0.3 (**)** | 156±42 (**) | 82±1.7 (****) |
| | $CH_3COONa$ | 100±40 (-) | 80±2.1 (**) | 147±22 (***) | 80±1.1 (****) |
| | NaCl | 104±27 (-) | 78±0.9 (**) | 98±8 (****) | 76±2.0 (****) |
| | PBS(LONZA) | 236±36 (**) | 81±1.4 (**) | 124±26 (**) | 78±0.8 (****) |
| | $H_2O$ | 76±39 | 66±3.0 | 14.5±1.8 | 40±1.9 |
| 4 weeks | $(NH_4)_2SO_4$ | 486±48 (****) | 84±0.7 (****) | 297±28 (****) | 78±0.3 (****) |
| | $Na_2SO_4$ | 170±32 (**) | 80±0.6 (****) | 338±79 (***) | 79±1.3 (****) |
| | $Na_2HPO_4$ | 126±16 (*) | **54±1.5 (****)** | 177±14 (****) | 77±0.5 (****) |
| | $CH_3COONa$ | 230±29 (***) | 79±1.1 (****) | 203±16 (****) | 76±0.6 (****) |
| | NaCl | 358±32 (****) | 81±0.5 (****) | 347±22 (****) | 78±0.8 (****) |
| | PBS(LONZA) | 189±33 (**) | 72±1.3 (***) | 206±27 (****) | 76±1.1 (****) |
| | $H_2O$ | 103±3 | 67±1.0 | 27±7 | 46±2.4 (****) |

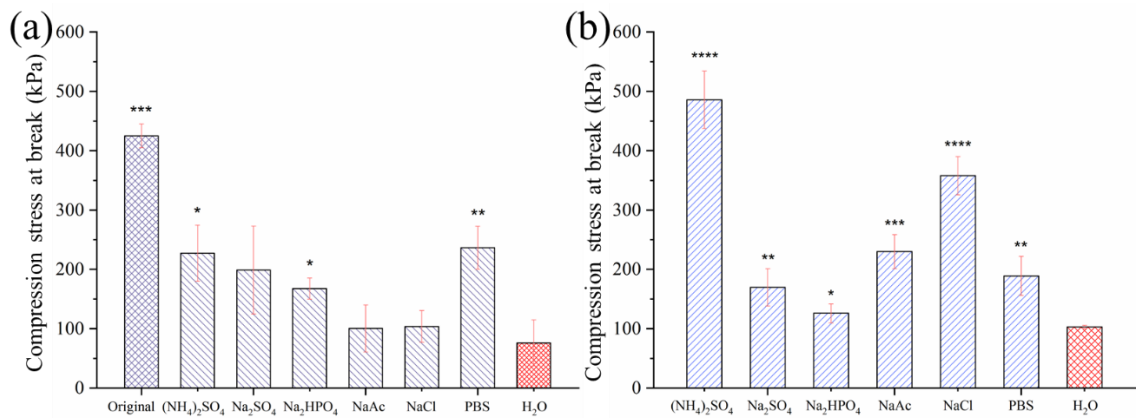

**Figure S3.** Compression stress at break of hydrogel C2-40 measured after synthesis ('Original') and after 1-day (a) and 4-week (b) incubation in different aqueous solutions. All the statistical analysis is presented with respect to $H_2O$ group and labelled as *$p < 0.05$, **$p < 0.01$, ***$p < 0.001$, ****$p < 0.0001$. All the data are presented as Mean±SD.

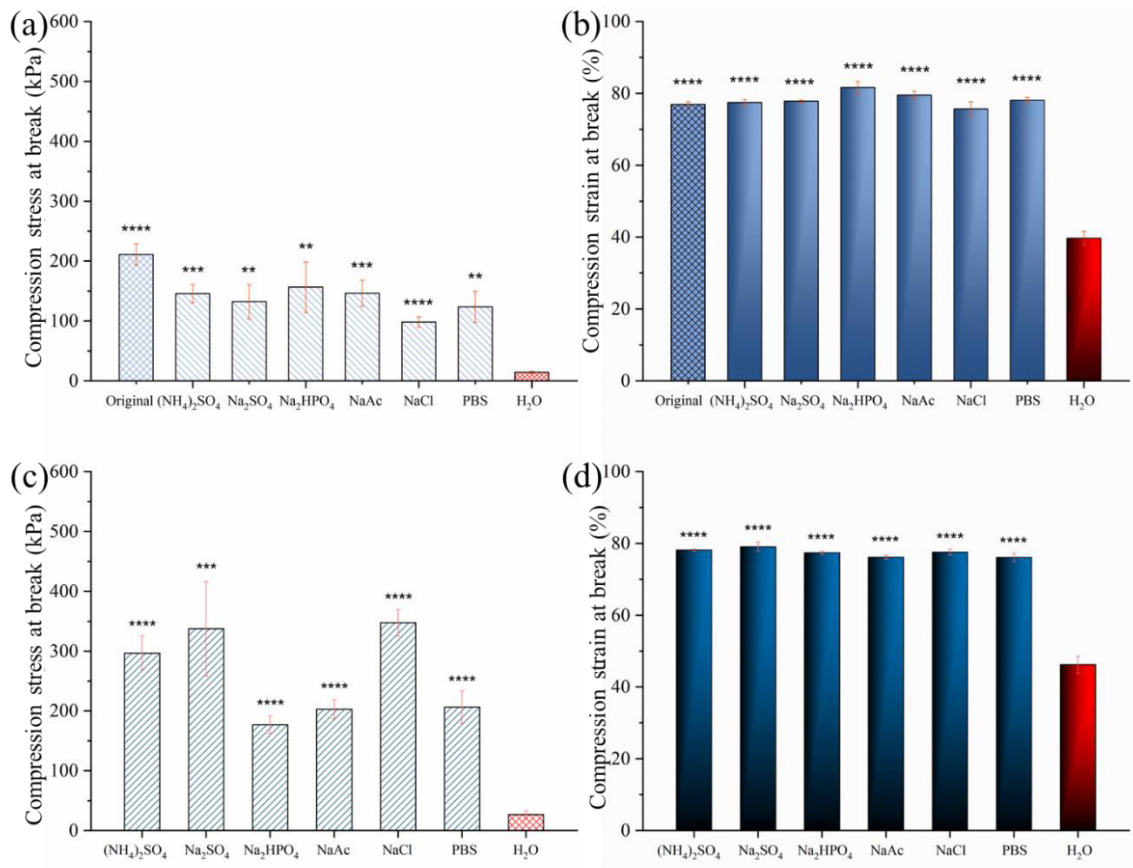

**Figure S4.** Compression measurements of hydrogel E2-40 after synthesis ('Original') and following 1-day (a, b) and 4-week (c, d) incubation in different aqueous solutions. All the statistical analysis is presented with respect to the H$_2$O group and labelled as **p < 0.01, ***p < 0.001, ****p < 0.0001. All the data are presented as Mean±SD.

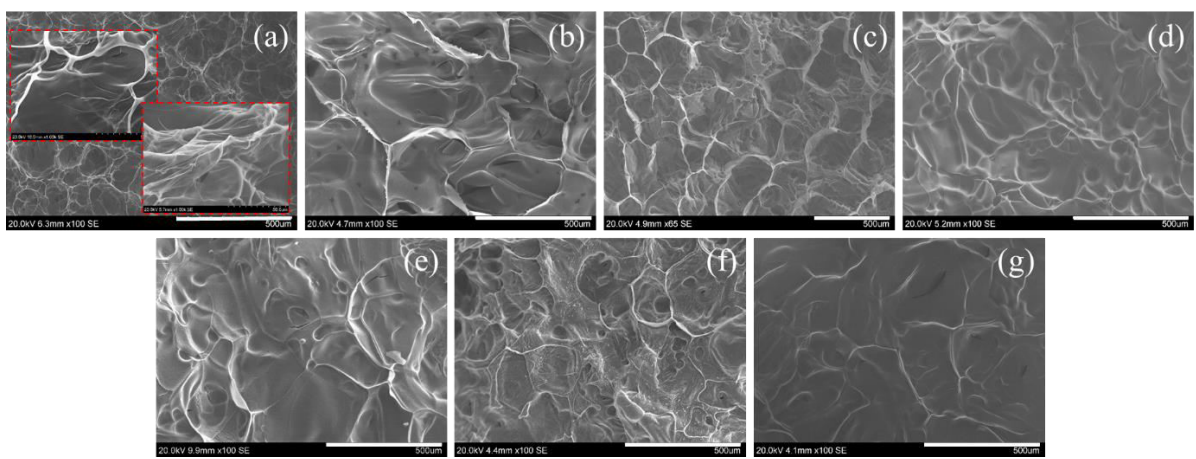

**Figure S5.** SEM images of freeze dried E2-40 networks following 4-week incubation in (NH$_4$)$_2$SO$_4$ (a), Na$_2$SO$_4$ (b), Na$_2$HPO$_4$ (c), NaAc (d), NaCl (e), PBS (f) and deionized water (g). Scale bar: 500 μm.

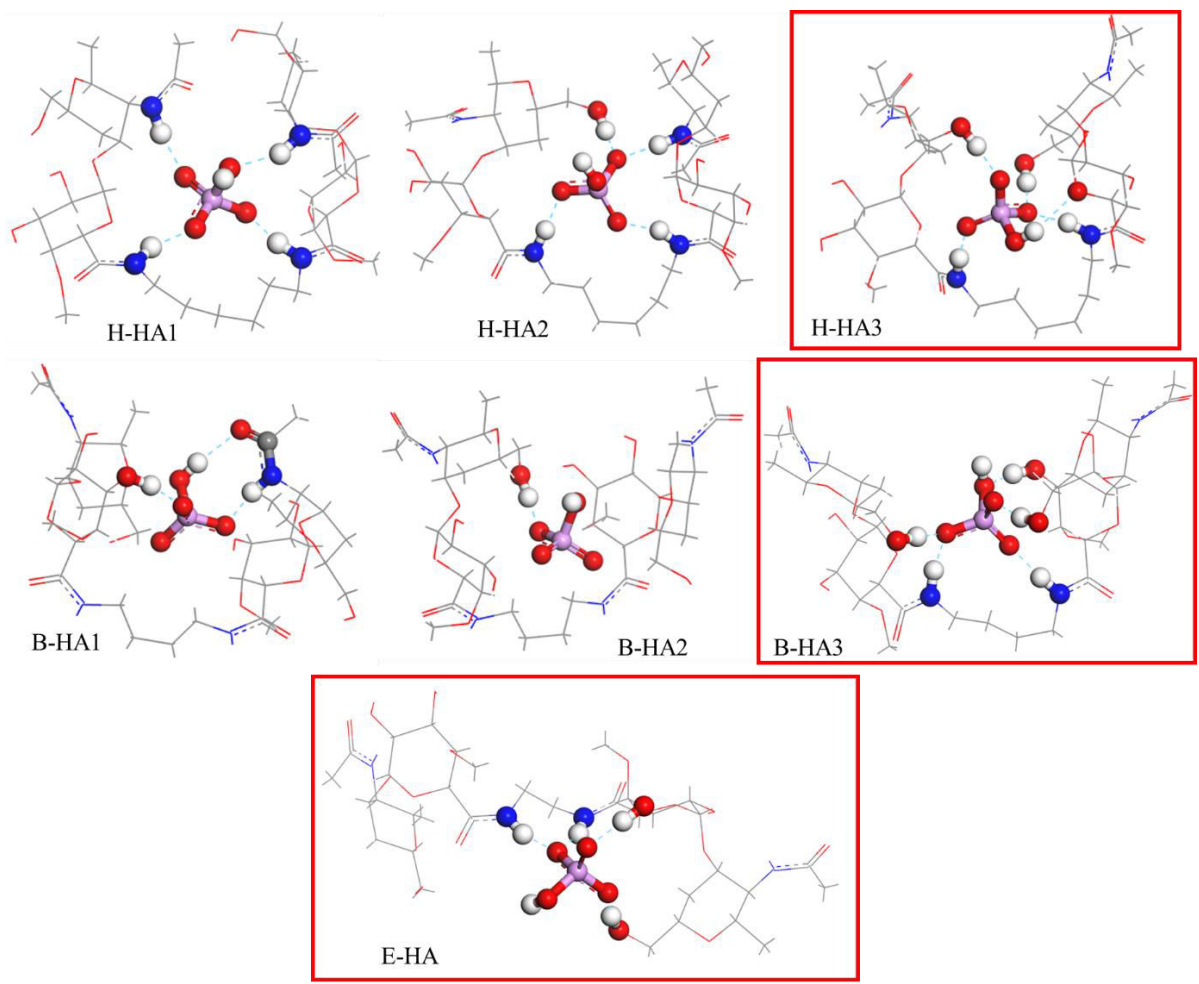

**Figure S6.** DFT calculations of the hydrogen bond interaction between $HPO_4^{2-}$ and hyaluronic acid crosslinked with either 1,6-Hexanediamine (H-HA), 1,4-butanediamine (B-HA) or Ethylenediamine (E-HA). In all models, oxygen (O) atoms were presented in red, nitrogen (N) in blue, sulfur (S) in yellow, carbon (C) in grey, hydrogen (H) in white and phosphorus (P) in pink. The optimised models with lowest interaction energy were highlighted by a red box.

**Table S2.** Optimised computing results of interaction energy in selected HA models.

| Model Name | ΔE (kcal/mol) |
|---|---|
| C-HA1 | -162.075 |
| C-HA2 | -169.501 |
| *C-HA3 | -170.751 |
| H-HA1 | -152.556 |
| H-HA2 | -160.341 |
| *H-HA3 | -162.149 |
| B-HA1 | -150.107 |
| B-HA2 | -146.913 |
| *B-HA3 | -167.491 |
| E-HA | -155.330 |

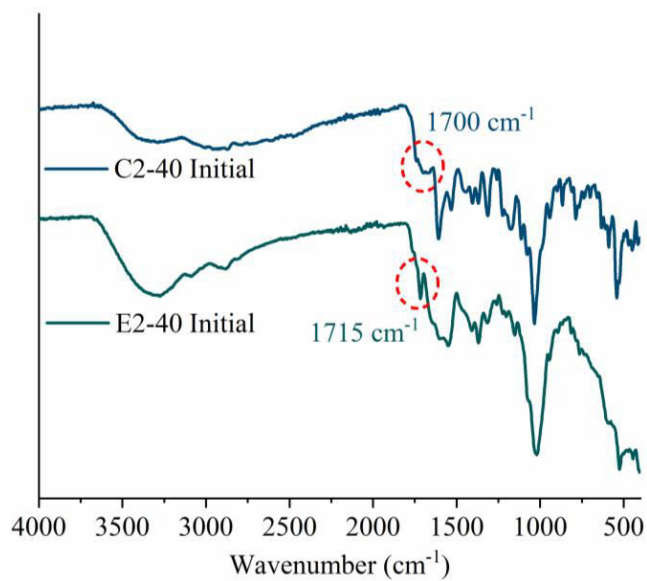

**Figure S7.** IR spectrum of freshly synthesised C2-40 and E2-40 networks.

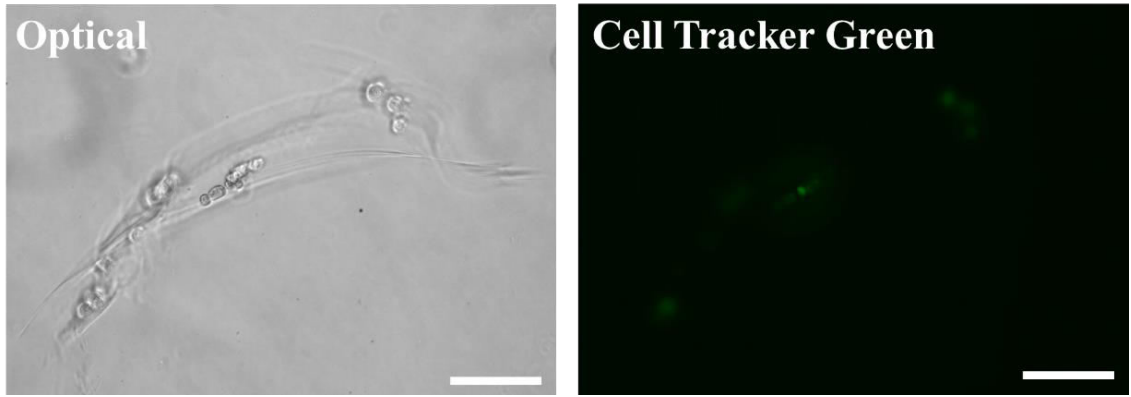

**Figure S8.** Cell Tracker Green labelled ATDC 5 cells on C2-40 surface after 2-day culture in basal medium. Optical image (left) and fluorescent image (right).

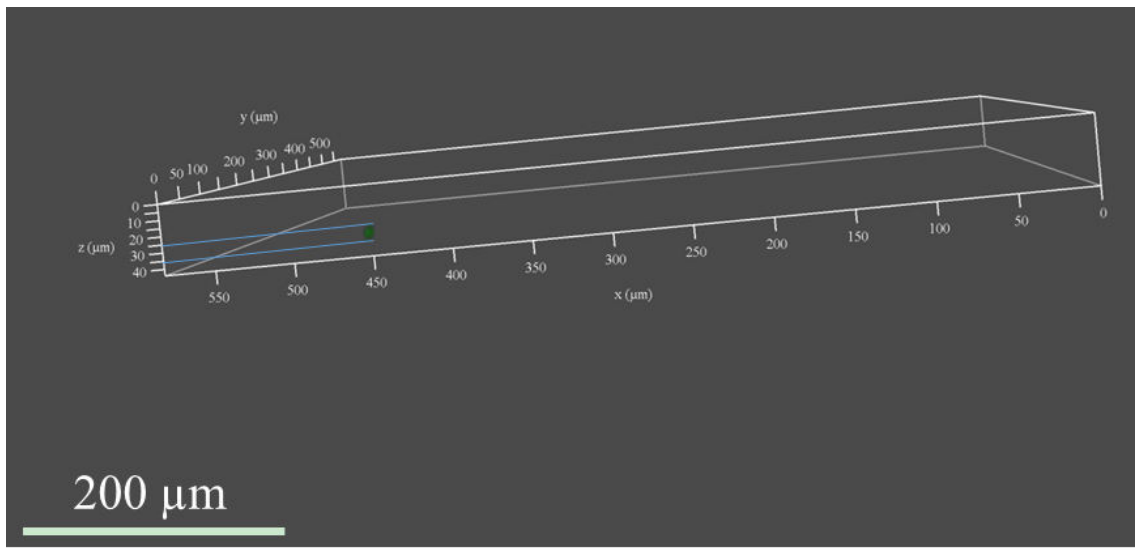

**Figure S9.** Laser confocal image of blank hydrogel control C2-40 in TM group following 3-day conditional culture.

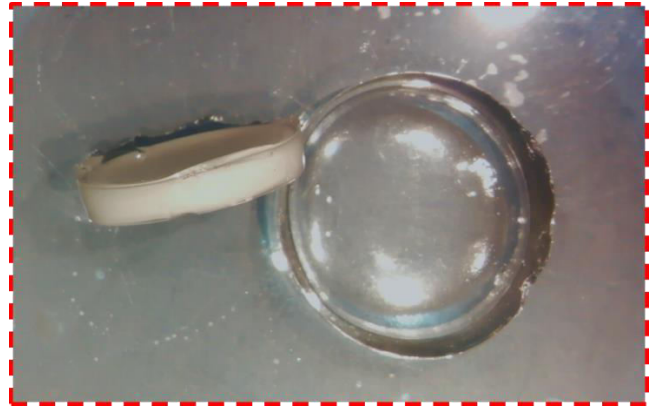

**Figure S10.** Optical images of the wet mineralized sample C2-40 following $HPO_4^{2-}$ treatment (side view, left) and C2-40 hydrogel control (top surface, right).

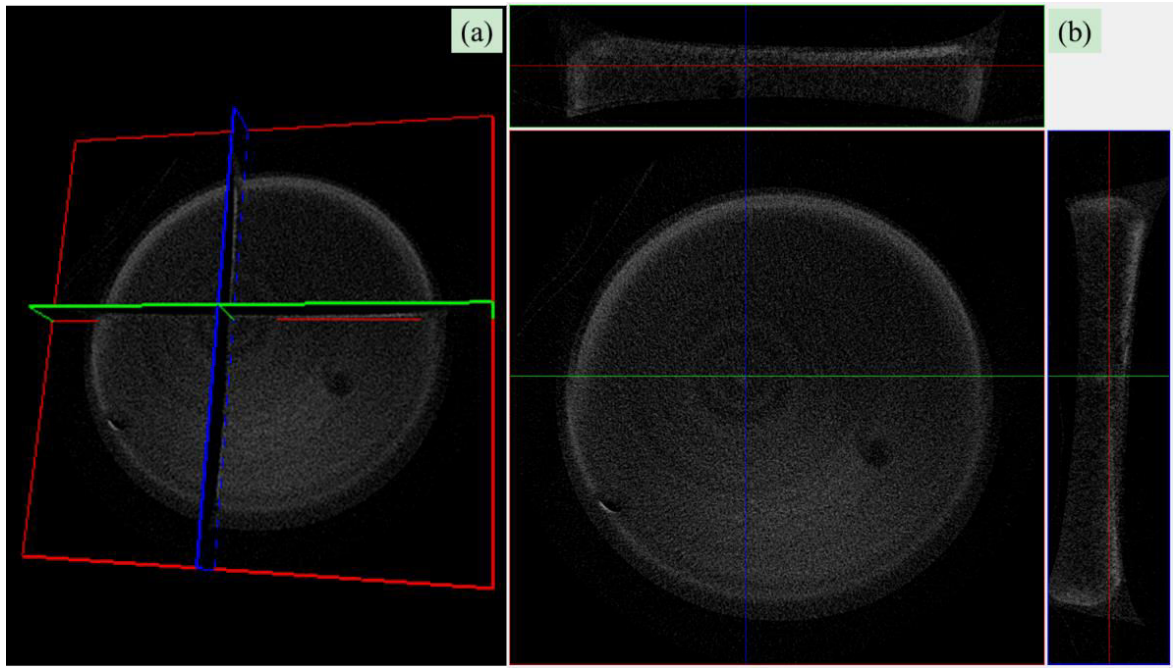

**Figure S11.** Macro-pattern study of mineralized sample C2-40 from μCT scan, combining 3-dimensional image (a) and cross section of each axis (b).

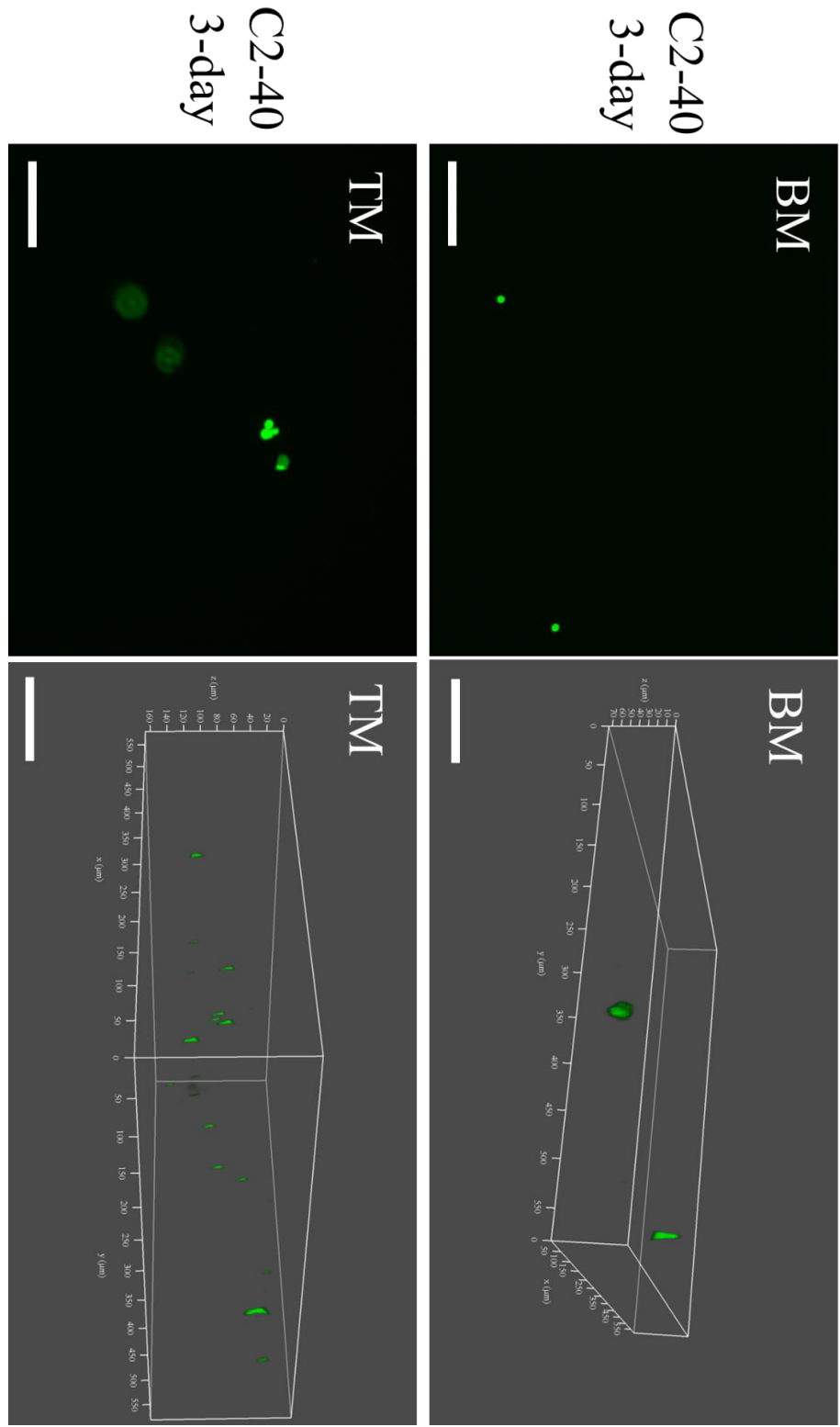

**Fig. S12.** First and second row of Figure 5 (in the main manuscript) in higher resolution. Conditional culture of ATDC 5 cell growth. Cells adhesion study on the surface of C2-40 hydrogels in either basal medium (BM) or $Na_2HPO_4$ treated medium (TM) after 3 days. Scale bar: 100 μm.

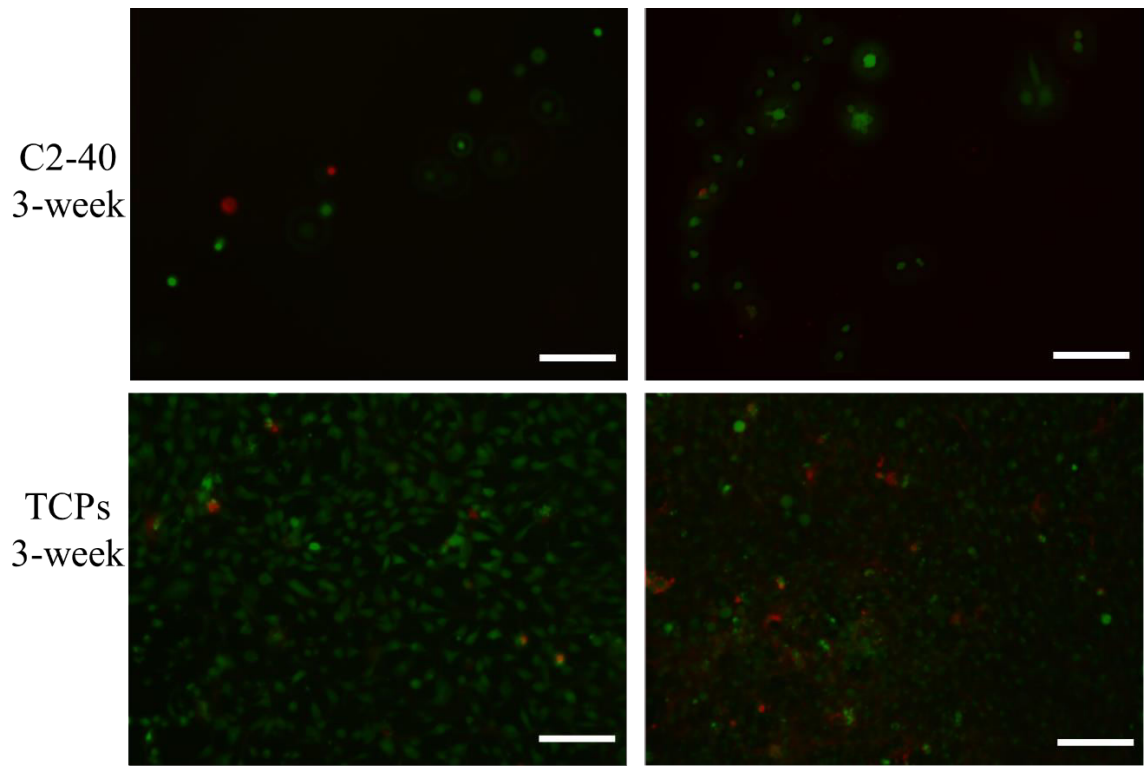

**Fig. S13.** Third and bottom row of Figure 5 (in the main manuscript) in higher resolution. Cells after 3-week conditional culture on C2-40 hydrogel surface (third row) and TCPs (bottom row) in either BM or TM group. Live labelling was presented in green and dead labelling was indicated in red. Scale bar: 100 μm.